# Exponential-growth prediction bias and compliance with safety measures in the times of COVID-19

Ritwik Banerjee[*,1], Joydeep Bhattacharya[2], and Priyama Majumdar[1]

[1]Indian Institute of Management Bangalore, India

[2]Iowa State University, USA

**Abstract**

We conduct a unique, Amazon MTurk-based global experiment to investigate the importance of an exponential-growth prediction bias (EGPB) in understanding why the COVID-19 outbreak has exploded. The scientific basis for our inquiry is the well-established fact that disease spread, especially in the initial stages, follows an exponential function meaning few positive cases can explode into a widespread pandemic if the disease is sufficiently transmittable. We define prediction bias as the systematic error arising from faulty prediction of the number of cases $x$-weeks hence when presented with $y$-weeks of prior, actual data on the same. Our design permits us to identify the root of this under-prediction as an EGPB arising from the general tendency to underestimate the speed at which exponential processes unfold. Our data reveals that the "degree of convexity" reflected in the predicted path of the disease is significantly and substantially lower than the actual path. The bias is significantly higher for respondents from countries at a later stage relative to those at an early stage of disease progression. We find that individuals who exhibit EGPB are also more likely to reveal markedly reduced compliance with the WHO-recommended safety measures, find general violations of safety protocols less alarming, and show greater faith in their government's actions. A simple behavioral nudge which shows prior data in terms of raw numbers, as opposed to a graph, *causally* reduces EGPB. Clear communication of risk via raw numbers could increase accuracy of risk perception, in turn facilitating compliance with suggested protective behaviors.

**Keywords: COVID-19, coronavirus, exponential growth bias, pandemic, prediction bias, WHO safety measures**

[*]This research has has received ethics approval from the Internal Review Board of Iowa State University. Funding from Indian Institute of Management Bangalore and Iowa State University are gratefully acknowledged. The paper presents early results from preliminary data collected as part of a larger project to understand behavioral response to COVID-19. Email: ritwikbanerjee@iimb.ac.in

# 1   Introduction

```
              "What on earth prompted you to take a hand in this?"
              "I don't know.  My...  my code of morals, perhaps."
               "Your code of morals.  What code, if I may ask?"
                                "Comprehension."

                          —— Albert Camus, The Plague
```

The COVID-19 outbreak is a global pandemic, adversely affecting the lives of millions of people around the world. While not a lot is yet known about the virus and how it operates, and with no vaccine or medication in sight, there is complete agreement on the need for people to strictly follow WHO guidelines regarding frequent washing of hands, use of hand sanitizers and face masks, social distancing and, if needed, self-quarantine (Lunn et al., 2020). Of course, unsurprisingly, not everyone complies with these guidelines, at least not with the seriousness with which they need to be followed (van Bavel et al., 2020; Cummins, 2020; Pinsker, 2020). This sort of (non)compliance is an active decision predicated upon, at least, one influential variable, the accuracy with which an individual perceives her own likelihood of getting infected (loosely, her risk perception).[1] All else same, if her risk perception is high, she is more likely to show compliance.

A large body of work in behavioral psychology and economics has documented how accuracy of risk perception may be compromised by a whole host of behavioral biases (such as, optimism, overconfidence, and so on). This paper focuses attention on one bias: exponential-growth prediction bias (EGPB), the "pervasive tendency to linearize exponential functions when assessing them intuitively" which leads to "a systematic tendency to underestimate a future value given a present value" (Stango and Zinman, 2009; Goda et al., 2019)[2]. Why is a study of this bias critical to our understanding of the current pandemic? Prior epidemiological studies (see Keeling and Rohani (2011)) document how disease spread, especially in the initial stages, often follows an exponential function. This means few positive cases can explode into a widespread pandemic if the disease is sufficiently contagious. It also means any failure to "read the tea leaves", arising from EGPB, can be devastating for society and, therefore, worthy of attention. Fig. A1 plots the trajectory of reported cases for four countries, Germany, the United States, France, and Spain. The growth trajectory is clearly exponential.[3] Pinsker (2020) argues "[t]he human brain can have trouble keeping pace with such rapid growth" and that "people tend to underestimate the speed at which exponential processes—such as a disease outbreak—unfold." This is our entry point. To the best of our knowledge, ours is the first attempt at understanding the role of the EGPB in the study of disease spread in a pandemic and its subsequent effect on compliance behavior.

---

[1] This point was noted in a classic article pertaining to the 1918 influenza pandemic in *Science* "People do not appreciate the risks they run" (Soper, 1919).

[2] Existing analysis of the exponential growth bias rely on quantifying the difficulty most people have with compound interest rates Levy and Tasoff (2016). General difficulty with discriminating linear from non-linear processes is documented in Cordes et al. (2019) and shows up even as early as in pre-kindergarten students (Ebersbach et al., 2010).

[3] In the U.S., the number of confirmed cases was 213 on March 9; six days later, it was 1,678. Over the same period, the number of confirmed cases in France rose from 1,116 to 4,469, and the number in Italy rose from 7,375 to 21,157—and Italy had only 1,689 cases as of March 2.

We define prediction bias as the systematic error arising from under or over -prediction of the number of COVID-19 positive detections $x$-weeks hence when presented with $y$-weeks of prior, actual data on the same. We call it EGPB if the actual data follows an exponential function and the predictions fail to appreciate the extent of the true convexity. Our analysis works off the premise that those who suffer from EGPB will greatly underestimate how quickly a disease spreads, fail to perceive their own onrushing risk, and hence, show low compliance with safety measures.

We use data from a global survey we conducted in March of 2020 to investigate three pressing questions of significant policy relevance:

1. How much of individual-level compliance with WHO guidelines can be explained by the bias associated with predicting the number of COVID-19 cases, after controlling for demographic and cultural variables?

2. Does EGPB diminish as a country moves through different stages of the disease? [4]

3. Does a simple nudge in terms of how data is presented help mitigate EGPB?

We ran an incentivized, data-collection survey on Amazon's Mechanical Turk, an online platform, with participation from people in 43 countries. The survey is not nationally representative, but since the samples all came from the same source, they permit a relatively clean comparison. Participants were also asked about their frequency of hand-washing and use of sanitizers and masks, their willingness to pay for masks, their view about the social appropriateness of others' behavior, their like/dislike of government responses; other demographic information was collected as well. Using this information, we generate composite indices measuring individual attitudes regarding their i) own compliance, ii) appropriateness of violation of WHO measures, and iii) satisfaction with the government's performance. Taken together, this gives us a broad sense of "compliance". The prediction data was used to construct several measures of EGPB and featured alongside the compliance measure as regressors in a multivariate regression model.

Our main results are as follows. First, we document the presence of EGPB as it pertains to forecasting the $x$-weeks ahead path of the disease. Second, the "degree of convexity" reflected in the predicted path of the disease is significantly and substantially lower than the actual path.[5] This connects with the first result: the source of the prediction bias is the "lower convexity" of the mental model used.[6] Third, EGPB is significantly lower for participants from countries at an early stage relative to those at a later stage of the disease. Fourth, we find our measures of EGPB are significant predictors of compliance: higher bias is associated with lower own safety compliance, a higher approval of violation of safety measures and a greater satisfaction with the government's response with respect to the pandemic. Fifth, we find that providing disease trajectory information using $y$-weeks prior data in the form of raw numbers *causally* reduces EGPB more than providing the same via a graph. Like us, Levy and Tasoff (2017) find "exponential-growth bias is unlikely to be eliminated by simple "nudges" such as a graphical intervention.

---

[4]We identify three distinct stages a country can be in: Stage 1 with less than 100 positive detections, Stage 2 with between 100 and 999, and finally, Stage 3 with 1000 or more.

[5]We use the term "degree of convexity" to mean the rate of change of the gradient of the data function. In our case, this rate is higher in the actual data function as compared to the predicted data function.

[6]Pennycook et al. (2020) find that overall cognitive sophistication (the composite of four measures) was a strong negative predictor of COVID-19 misperceptions. To the extent discriminating between linear and exponential processes is correlated with cognitive sophistication, our results are in sync with their findings. Strikingly though, their measure of cognitive sophistication was not a strong or consistent predictor of COVID-19 risk perceptions or behavior change intentions. Similarly, Stanley et al. (2020) find that "individuals less willing to engage effortful, deliberative, and reflective cognitive processes were [...] less likely to have recently engaged in social-distancing and hand-washing."

There is a small yet burgeoning literature studying the role of biases in the context of the COVID-19 outbreak. Our paper was written concurrently and independently of this literature. Our paper is closest in spirit to an important contribution by Wise et al. (2020) who, like us, use an online sample (but only of U.S. households) between March 11, the day when the WHO declared COVID-19 a pandemic, and March 16. Their primary focus is on deciphering to what extent individuals are aware of their own risk of contracting the disease, their risk of passing it on to others, and the extent to which protective behaviors are predicted by their perception of risk. Our study is different in two ways. First, we do not directly elicit peoples' perception of their own risk, because as Wise et al. (2020) report, such elicitations may be contaminated by the simultaneous presence of multiple cognitive biases (such as optimism or overconfidence; see also Stanley et al. (2020)).[7] Second, our measure of perception bias, EGPB, is rooted in the linear vs exponential heuristic literature with a clear theoretical underpinning. And third, our measures of prediction bias are less likely to be "pure noise" since they are the result of an incentivized elicitation process.

A portion of the major contribution by Fetzer et al. (2020) is also devoted to unearthing the sort of prediction bias that is the focus of our paper. In their setup, online participants are asked to predict the number of infections from a fictitious disease over several days under several formulaic scenarios designed to test their innate ability to see geometric or exponential growth patterns. They find that the predictions of the median participant could be well approximated by a linear, mental model. We depart from their work in several ways. First, our design is not aimed at understanding whether participants can "do the math" and figure out where a hypothetical series is headed. Second, we provide actual, COVID data on the number of *reported* cases (not infections, since in the absence of 100% testing, it is impossible to know the head-count of infections; if anything, participants likely form priors about infections based on the number of reported cases, and such priors could depend on their innate faith in data-reporting bodies, which, by the way, we test for). Third, our results show our set of participants do not seem to have a linear model in mind: their mental model is neither a linear extension of best fit over the entire data range nor is it a linear extrapolation from the last shown data point.

---

[7]That is not to say these other biases cannot play a role; it is just that for a disease like COVID, the EGPB is of first-order importance. Also, our measures of EGPB are composite in nature. While composite, summary measures are useful to policymakers, the benefit of the granular measures in Wise et al. (2020), capturing own risk of getting infected versus the risk of infecting others, is also high.

## 2 Experimental Design and Procedure

We show our participants data on the actual number of COVID-19 (same as in Fig. A1) from four countries majorly affected by the virus as of 21st March, 2020, namely, Germany, USA, France, and Spain. In our experiment, participants perform four prediction tasks for two out of these four randomly chosen countries; country names are not revealed.[8] In each task, they are shown three, actual weekly data points of COVID-19 cases. Subsequently, they are asked to predict the number of cases for Weeks 4 and 5, the actual numbers of which is known to us, the researchers. This means, we, the researchers, have full knowledge of the true, underlying data-generating process.

Formally, denote the actual and the predicted number of COVID-19 cases in Week $i$ by $N_i$ and $P_i$, respectively, for $i = 1, 2, 3, 4, 5$. As noted earlier, the participants observe three data points on the number of COVID-19 cases in three consecutive weeks $(N_1, N_2, N_3)$ and are asked to make their predictions for Week 4 and Week 5.[9] The bias for Week 4 is defined as the difference between the actual number ($N_4$) and the predicted number in Week 4 ($P_4$), relative to the difference between the actual number in Week 4 ($N_4$) and the actual number in Week 3 ($N_3$). In other words, the actual prediction error relative to the maximum possible error in Week 4 may be interpreted as the Bias for Week 4 with respect to Week 3 i.e., $Bias_{43} = \frac{N_4 - P_4}{N_4 - N_3}$.[10] Similarly, we define Bias for Week 5 with respect to Week 3 as $Bias_{53} = \frac{N_5 - P_5}{N_5 - N_3}$. Finally, we analyze the results in terms of $Bias_{avg}$, which is the average of $Bias_{43}$ and $Bias_{53}$.

Participants are paid $0.50 as a participation fee. Further, two of the four prediction tasks are randomly picked and if the participant's prediction is within 5% of the actual number, she is paid an additional $0.25 for that task. Participants can earn a maximum of $1 for the entire experiment, which lasted for about 7 minutes.

Notice, from Fig. A1, that the early phases of the spread of COVID-19 is described by notably less convexity relative to later phases. In order to understand, whether the prediction accuracy varies with respect to the phases of the disease spread, we implement a within-subject variation in the four predictions tasks. In two consecutive tasks, the participants are shown the data either from the early phases or from later phases of the two randomly selected countries. Asking them to respond to early and later phases ensure that their personal experience in their home country is mimicked in at least one of the tasks. If in the first two tasks, a participant is shown data from two countries at early phases, in the next two they are shown the numbers from the later phases of the same two countries. The order in which a subject sees the early phase is randomized to control for order effects.

We implement this design so that we can explore whether the nature of representation (graphical or numeric) of the actual data helps mitigate EGPB. To that end, some participants are randomly shown the actual number of COVID-19 cases graphically (as is the dominant form of representation of the data in print or online media) while others are shown the same in terms of raw numbers. We implement this through a between-subject design, which allows us to estimate the *causal* effect of the nudge in mitigating the EGPB. Screen 5 of the experimental instruction given in Appendix 4 presents an example of the two forms of data representation .

---

[8] If a participant belongs to any of the four countries, say X, she is not be shown numbers from X; instead she is shown numbers from two countries randomly chosen from the set W, Y and Z. This ensures her prior information about disease progression in her own country, X, does not contaminate her prediction.

[9] Since the true number of infected individuals at any point is unknown, we go by the official statistics on reported cases. Interestingly, respondents in our sample report that their belief about true infection rate is, on average, 10% higher than the official statistics.

[10] We rely on the simple deviation instead of root mean square deviation since the latter will transform each deviation into a positive number, which in turn, will lead to overweighing the deviation.

We define Stage 1 as countries as those with less than 100 positive cases, Stage 2 between 100 and 999, and Stage 3 with 1000 or more cases, as of 21st March, 2020. Running the study on Amazon MTurk got us access to a global pool of participants who resided in countries that were themselves at different stages of the disease. This gives us an unique opportunity to study whether the EGPB varies with the stage of the country in which a participant resides.

Following this, we administer a short survey to capture the participants' own protection behavior and compliance with the WHO guidelines. The survey, along with the experimental instructions, is included in Appendix 2. Participants are asked to report their frequency of hand-washing, use of sanitizers and masks, their opinion about the appropriateness of public gatherings and the like, their perception about the desirability and efficacy of their own government's responses. Basic demographic information was also collected. We categorize the questions into three broad headings, namely, Actual Realized Compliance, Appropriateness of Violation of Safety Norms and Agreeableness with Government Performance, which taken together offers a comprehensive sense of participants' compliance with safety guidelines. The definitions are given in Table A1. In addition, we randomize the order of whether a participant is shown the prediction segment first or the survey segment first in order to mitigate experimenter demand effects.

In sum, the experimental design allows us to answer the three main research questions in this paper: 1) Do people exhibit EGPB in the context of the COVID-19 spread? 2) Is this bias a significant predictor of lax attitude towards COVID-19 related safety norms? 3) Can a simple nudge in the way the data is represented help mitigate the bias?

## 3 Results

Our first set of results document existence of the aforediscussed prediction biases. Fig. 1A presents the number of COVID-19 cases in early and later phases, for all the five weeks along with the mean, median and the interquartile range of the predicted number of cases in Weeks 4 and 5, averaged over all the four countries. Fig. 1B presents the same in terms of the log-transformed data[11]. The mean prediction in the early phase panel of Fig. 1 exceeds the actual number because of outliers. Except for the one case, both the median and the mean prediction lies well below the actual number of COVID-19 cases. To avoid outlier-driven distortions, we rely on the median measure in all subsequent analysis. The actual numbers, averaged across all four countries, are 1540 in Week 4 and 9189

---

[11] We present the country-specific predictions on the log transformed data for each of the four countries separately in Fig. A2

in Week 5 for the early phase; the respective medians are 521 and 1081, respectively. The corresponding numbers in Week 4 and Week 5 are 17450 and 55934, while the median predictions are 9500 and 18000, respectively.

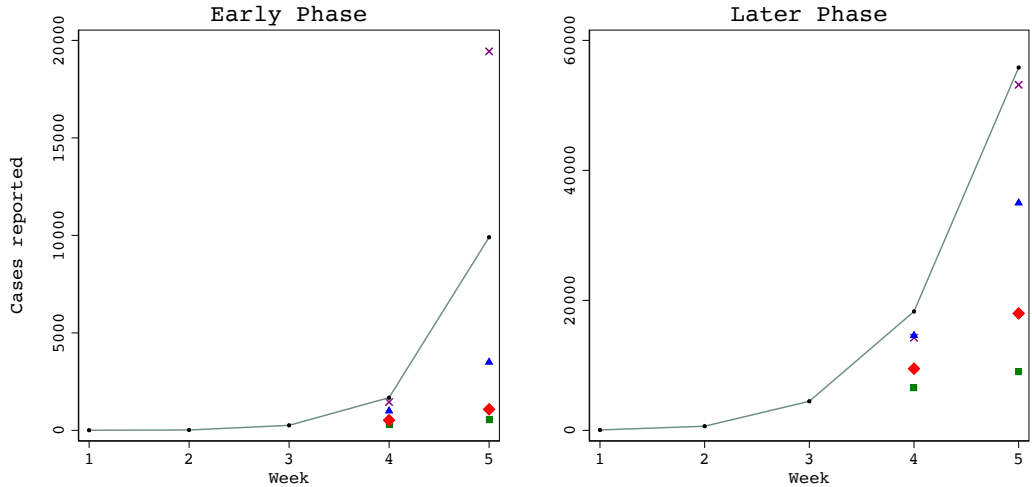

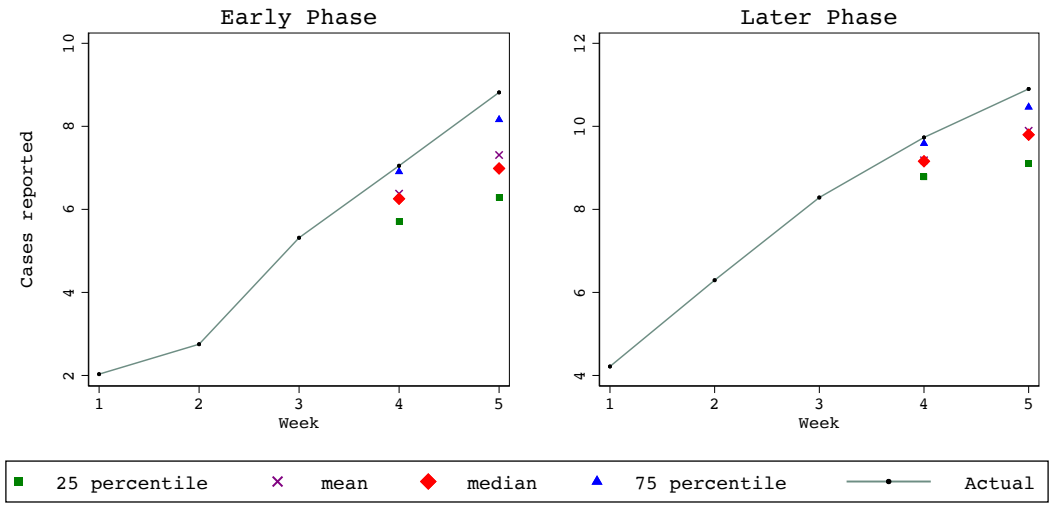

Figure 1: Actual and predicted number of COVID-19 cases pooled across four countries

*Panel A above plots the actual number and the predictions of COVID-19 cases in the linear scale, while Panel B plots the same data in a logarithmic scale. Both the panels use data pooled across the four countries. The predictions for early (later) phase cases are given on the left (right). Each graph presents participants' 25th percentile, mean, median, and 75th percentile prediction of the number of COVID-19 cases on Week 4 and Week 5.*

As discussed earlier and shown in Fig. A1, the underlying data generating process of actual spread of COVID-19 is convex. One interesting question is, is the prediction model used by the median individual also convex or is it linear? To test this, we compute the ratio of the slopes of the line segment connecting $N_4$ and $N_5$ and that connecting $N_3$ and $N_4$ (i.e., $\eta = \frac{slope(N_4,N_5)}{slope(N_3,N_4)}$). We compare $\eta$ with the ratio of slopes of the line segment connecting $P_4$ and $P_5$ and that connecting $N_3$ and $P_4$ (i.e., $\rho = \frac{slope(P_4,P_5)}{slope(N_3,P_4)}$). Column (3) in Table 4 reports $\eta - \rho$. We test $H_o : \eta - \rho = 0$ and show it is significantly different from zero for all the countries, for both phases (except in one case). This indicates that the underlying prediction model used by the median individual is significantly less convex than the actual data generating process; however, we do not find evidence that the prediction model used is actually linear even though it is clear that the prediction model is significantly closer to the linear fit than the exponential.[12]

How does the degree of convexity of the predictions vary with the stages of the disease a participant is exposed to? We find that participants from Stage 3 countries (relative to those from Stages 1 and 2), for all our bias measures, make predictions that are closer to the best-fit linear model while predictions of those from Stage 1 countries are closer to the actual, exponential data. This has the prima facie implication that people in advanced stages of the disease outbreak may perceive less, not more, convexity.

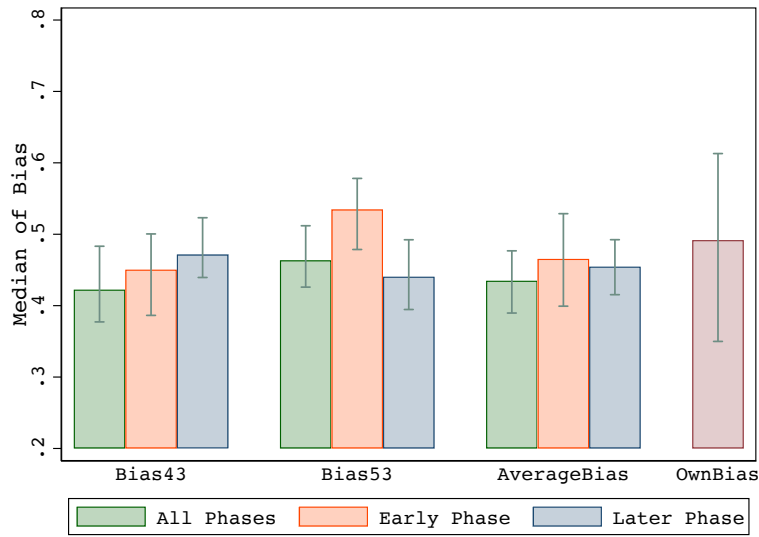

Figure 2: Prediction Bias

*Note: The figure reports the median values of different measures of biases. The error bars represent the 95% confidence intervals from a Kendall's tau test for the hypothesis that the median is zero.*

We carry out the rest of the analysis on log transformed data with the median as the primary statistic of analysis. In Fig. 2, we analyze whether these differences, when transformed as biases, are significantly different from zero. $MedianBias_{43}$ is positive and significantly different from zero at 0.42 (Wilcoxon signed rank test, $p-value < 0.01$), meaning the median participant exhibits 42% under-prediction. Similarly, $Bias_{53}$ and the $AverageBias$ are positive and significant at 0.46 and 0.44 (Wilcoxon signed rank tests, $p - value < 0.01$). Clearly, the participants exhibit substantial prediction bias. Interestingly, $Bias_{43}$ is significantly smaller than $Bias_{53}$ (nonparametric median test, $p-value < 0.01$), suggesting that the size of the prediction error increases with time. While the main focus of the paper is on EGPB arising from the incentivized prediction, we also elicit participants' belief about the number of COVID-19 cases on the day of the experiment in their own country and their prediction about the same seven days hence. As Fig. 2 indicates participants' (non-incentivized) prediction about the number of cases in their own country, seven days hence, reveals a 49% under-prediction.

---

[12]We test if the prediction model used by our participants is indeed linear by comparing the slopes of the individual linear pieces connecting $N_3, P_4, P_5$ with the slope of the best linear fit of $N_1, N_2, N_3$ ($\hat{\beta}$). The nonparametric equality of median test rejects the null that the piece wise slopes $N_3, P_4$ and $P_4, P_5$ are in fact equal to the slope of the linear fit. Further, we statistically compare the euclidian distance between the median prediction and the linear fit, and the median prediction and the actual data. We find the prediction model to be significantly closer to the linear fit. These results are not reported but are available upon request.

Our Amazon-MTurk participant pool comprised of people from 43 countries, each country being at different stages of COVID-19 spread at the point in time when the data was collected. This gives us an unique opportunity to examine how EGPB compare for people from countries at different stages of the disease. Fig. 3 compares the biases across Stage 1, Stage 2 and Stage 3 and plots the regression coefficient of Stage 2 dummy in Fig. 3(i) and (ii). While the incentivized bias measures are not different between Stage 2 and Stage 1, the $OwnBias$ is significantly higher in Stage 2 than in Stage 1 ($p-value = 0.02$). Fig. 3(iii) and (iv) shows that $Bias_{43}, Bias_{53}, AverageBias$ and $OwnBias$ are significantly higher in Stage 3 than in Stage 1 ($p-value < 0.5$ for all the variables). The pattern is less clear when Stage 3 is compared with Stage 2 as Fig. 3(v) and (vi) reveals. In all, the data suggests, biases, as per our different measures, are significantly higher for countries at Stage 3 than those in Stage 1 or 2.[13] Table 4 reports estimates and the standard errors.

---

[13]Relatedly, participants from Stage 3 countries use a prediction model that is closer to the best fit linear model than the true exponential data generating process, relative to those from Stages 1 and 2. The results not reported but are available upon request.

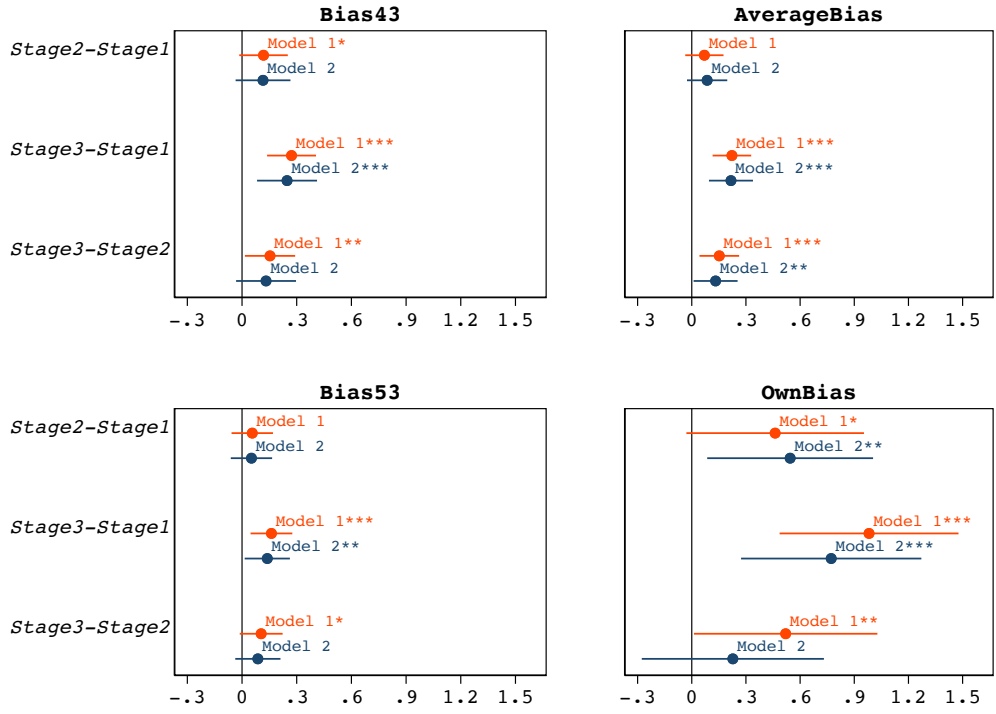

Figure 3: Variation of prediction bias between countries at different stages of COVID-19 spread

*This figure plots the differences in the median EGPB across countries in Stages 1, 2 and 3 of COVID-19 spread for each of the four measures of bias. Model 1 (Model 2) shows the pairwise differences in EGPB between the three stages estimated from a median regression without (with) controls. The control variables include age, gender, health condition, education level, income, treatment and log of reported COVID-19 cases as on 21st March (fixed for each country). The specification in Model 2 for OwnBias in (iv), additionally controls for an individual's information bias. * $p < 0.10$, ** $p < 0.05$, *** $p < 0.01$.*

Having established a significant EGPB, we ask, is it a significant predictor of compliance with safety measures. To examine this, we regress different indices of compliance on our EGPB measures, where the indices are computed from answers to a menu of questions asked of the participants[14]. We categorize the questions into three indices: Actual Realized Compliance, Appropriateness of Violation of Safety Norms and Agreeableness with Government Performance. The indices are constructed by taking the first principal component of the relevant set of variables. Table A1 lays out the summary statistics of the compliance indices. Each column of Fig. 4 corresponds to each of the above indices and plots the estimated regression coefficient for each definition of prediction bias. Model 1 and 2 run the regression without control variables and with control variables, respectively. As the figure illustrates and Table 4 confirms, EGPB is a negative predictor of Actual Realized Compliance, indicating higher the bias, lower is the self-reported measure of compliance of safety norms. A higher EGPB also predicts a higher Appropriateness of Violation of Safety Norms as the plotted coefficients in the second column of Fig. 4 reveals. This means, an individual who shows EGPB is also likely to view violations of safety norms such as the avoidance of public gatherings, working from home, etc. as not very alarming. Finally, EGPB is also a significant predictor of Agreeableness with Government Performance, implying, higher bias is associated with greater satisfaction with the performance of the government with regard to the measures taken *vis-a-vis* COVID-19.

---

[14]Appendix 4 presents the entire experimental protocol, along with the survey questions used to construct the indices.

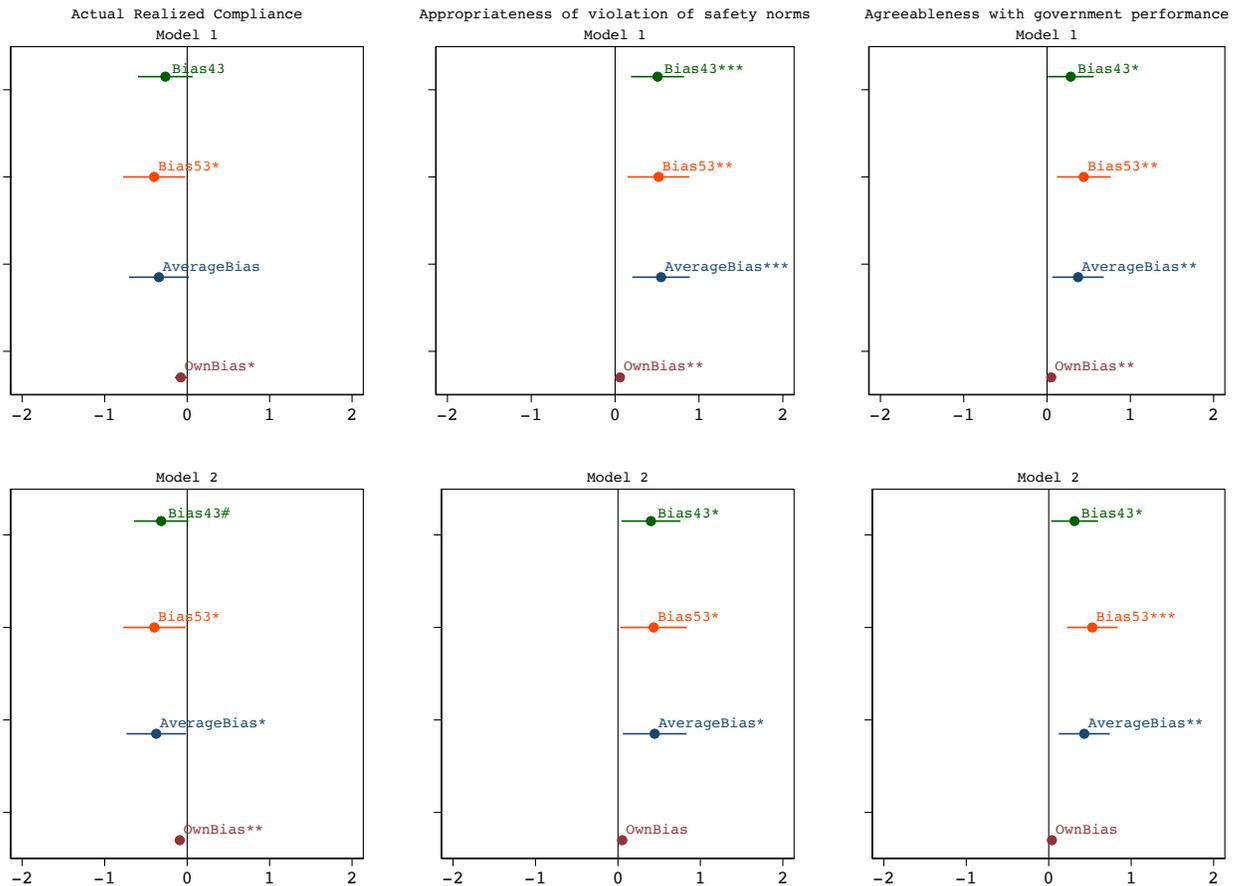

Figure 4: EGPB and COVID-19 compliance

*This figure plots the coefficients estimated from OLS regressions between EGPB and compliance, measured via three indices: Actual realized compliance, Appropriateness of violation of safety norms and Agreeableness with government performance. Model 1 (Model 2) shows the estimates from specification without (with) controls. The control variables include age, gender, health condition, education level, income, perceived effectiveness of the safety measures, treatment and log of reported COVID-19 cases as on 21st March (fixed for each country). The specification in Model 2 for OwnBias in (iv), additionally controls for an individual's information bias. # p=0.11, *p < 0.10, **p < 0.05, ***p < 0.01.*

Our analysis thus far suggests that people predict using a model which is substantially less convex than the true data generating process. This generates significant prediction bias which, in turn, is significantly associated with non-compliance of safety measures. Given this link, we ask: would a simple perceptual nudge help ameliorate the prediction bias? To that end, we use a randomized experiment design in which some participants are randomly shown $N_1, N_2, N_3$, in terms of raw numbers, while others are shown the same in graphical form.

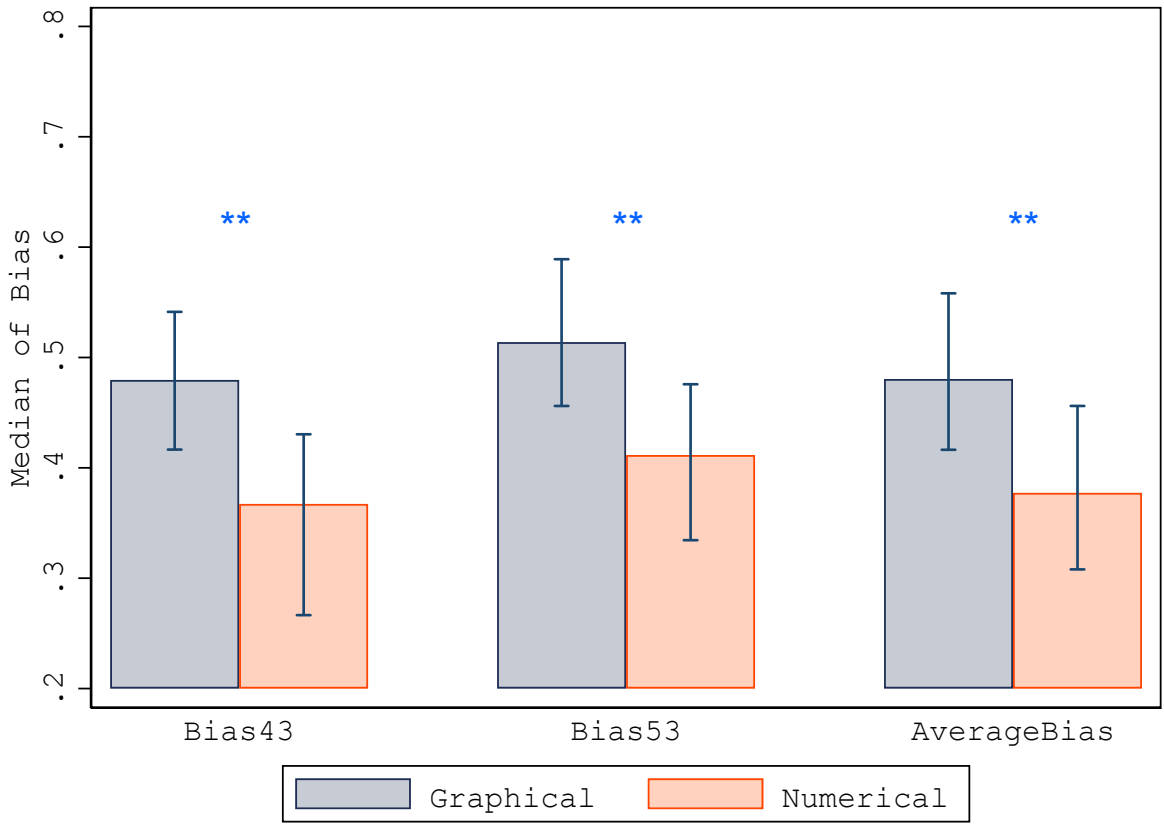

Figure 5: Prediction bias across treatments

Participants are randomized into graphical and numerical treatments, where they are shown the data from Weeks 1, 2 and 3 in the form of raw numbers and graphs, respectively. The figure plots the median prediction biases for the graphical and the numerical treatments. The error bars represent the 95% confidence intervals from a Kendall's Tau test for the hypothesis that the median is zero. * p<0.10, ** p<0.05, *** p<0.01.

The median biases across graphical and numerical treatments are presented in Figure 5. Presentation of the past data in numerical form significantly decreases the bias, however defined, relative to the graphical treatment. The treatment effects from ordinary least square regressions are reported in Table 4. This result holds with and without a set of controls. In fact, the treatment effect is negative and significant at the 1% level when the regression controls for a set of demographic variables. The regressions when run separately for the early and later phases reveal the same pattern. We do not find any heterogeneity in the additional robustness checks we run to see if the treatment effect varies with education levels and age[15].

---

[15]The results related to heterogeneity with respect to age, education levels and phases are not reported in the paper, but are available on request.

# 4  Conclusion

This paper uses data from a global survey to ask if perception errors relating to the tendency of people to linearize essentially exponential disease progression is key to understanding their compliance with widely recommended safety precautions. Our principal findings are a) there is evidence of rampant prediction biases; b) the mental model used to make predictions is not linear but substantially less convex than the actual; c) higher prediction bias is a significant predictor of lower safety compliance, less disagreement with violations of safety measures, and a greater satisfaction with the government's response with respect to the pandemic. Numerical representation of the disease trajectory can causally reduce prediction bias as opposed to graphical representation of the same. This last point suggests that raw numbers continue to make quite an impact on people's risk perception and should, at the very least, be presented alongside the common "flatten the curve" style graphics.

A few caveats are in order. First, there are well-understood problems (representativeness and external validity) of conducting global surveys on Amazon's MTurk that are germane to our survey. In our defense, though, these problems are also present in any of the online surveys that researchers are forced to use while the pandemic rages on. One key difference with existing work is our use of incentivized prediction elicitation. Another is that unlike existing computations of prediction bias Fetzer et al. (2020) using fictitious information, we show actual data from COVID infections and detect prediction bias, not just a mathematical inability to compute exponential progression.[16] Also, the timing of our survey was crucial: we wanted to know if the prediction biases changed in real time as a country moved from under a hundred infections to over a thousand infections. By now, most countries have moved on to Stage 3 but our data on biases in earlier stages may be useful for overall understanding of future epidemics.

Our work was primarily designed to detect prediction biases and to see if such biases were significant predictors of self-protection behavior. We are, however, unable to make definitive causal statements connecting prediction bias with compliance. Nor are we able to offer causal mechanisms to explain our findings. In particular, it is a bit perplexing why participants from countries at a later stage of the disease show larger prediction biases and less "convexity". One can speculate this has something to do with hope or optimism or widespread messaging related to "the end is near". We are able to causally show that messaging with raw data is more helpful in reducing the prediction bias than graphical data. Future work should take these exercises up.

---

[16] In Fetzer et al. (2020), "participants were instructed to assume that on day 1, one person has the fictitious disease. Furthermore, they were told to assume that each day a newly infected person infects two healthy people and then stops being contagious. Participants were further told that on day 2, 3 people will be infected by the disease as the person who had the disease on day 1 spread it to two other people on day 2. Participants were then asked to predict the count of total people infected with the fictitious disease on day 5, 10, and 20."

# Appendix 1

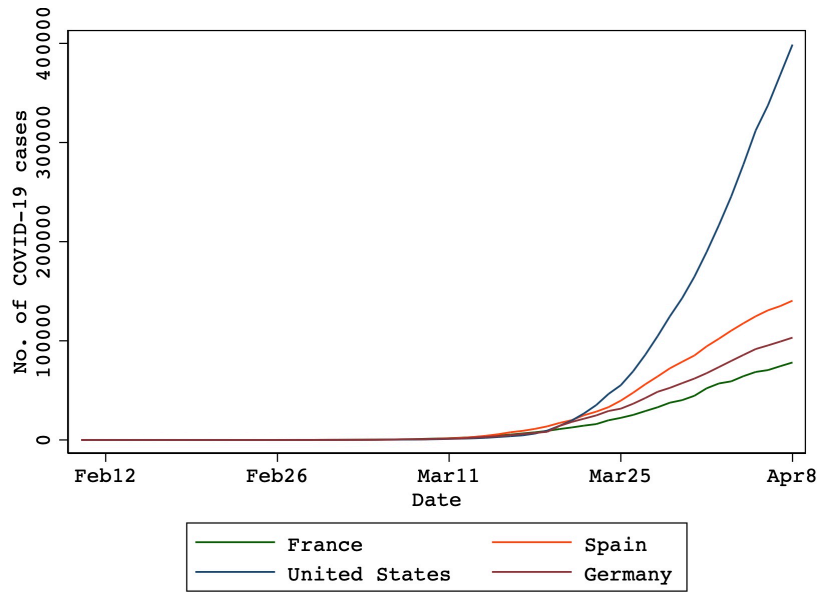

Figure A1: COVID-19 cases in Germany, US, France, Spain

The figure plots the growth trajectory of COVID-19 positive cases in Germany, US, France and Spain between February 10, 2020 and April 8th, 2020. It is clear that the growth path of the virus infection is exponential in nature.

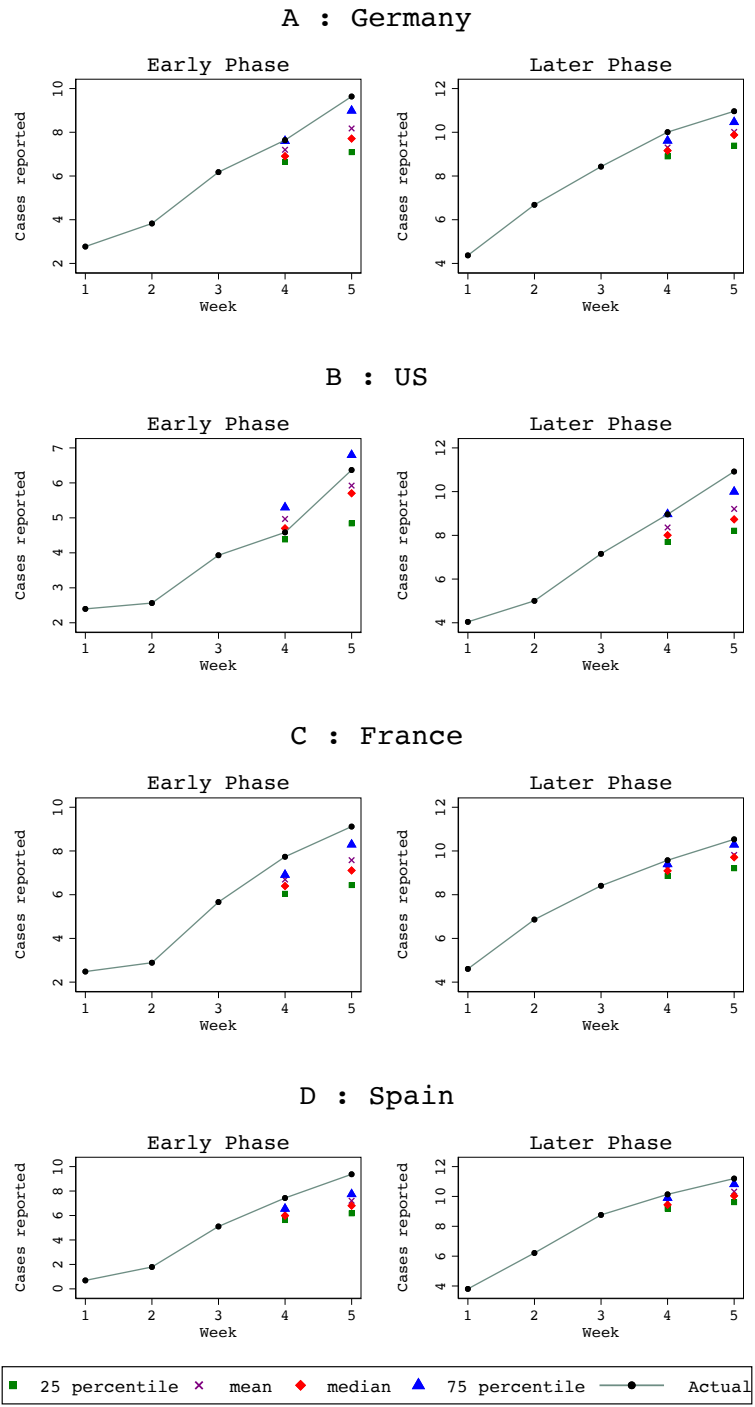

Figure A2: Country wise prediction bias in logarithmic scale

The figure plots the five actual number of COVID-19 cases, mean, median and interquartile range of the predictions, on the logarithmic scale. Participants observe the actual data in Week 1, 2 and 3 and predict the number of cases in Week 4 and 5. The line connecting the actual number of cases give the growth path of the disease. The panels on the left (right) plot the above for the early (later) phases of the disease transmission. In all the cases above, except USA in the early phase, the mean and median are significantly less than the actual number. The mental model used by the participants is significantly less convex than the actual data, as is clear from the growth path of the median of predicted number of cases.

Table A1: Summary Statistics

| Variable | Definition | Median | Mean Absolute Deviation from Median |
|---|---|---|---|
| Bias43 | Difference between the log of actual and predicted number in Week 4, relative to the change in log of actual number of COVID-19 cases between Week 3 and 4 $Bias43 = \frac{N_4 - P_4}{N_4 - N_3}$ | 0.42 | 0.35 |
| Bias53 | Difference between the log of actual and predicted number in Week 5, relative to the change in log of actual number of COVID-19 cases between Week 4 and Week 5 $Bias53 = \frac{N_5 - P_5}{N_5 - N_3}$ | 0.46 | 0.29 |
| AverageBias | Average of $Bias43$ and $Bias53$ | 0.43 | 0.31 |
| OwnBias | Difference between the log of actual and predicted number of COVID-19 cases one week later in one's own country | 0.49 | 1.87 |
| Own country information bias | Difference between the log of actual and perceived number of COVID-19 cases on the day of response in one's own country | 0.001 | 0.10 |

| Variable | Definition | Mean | Standard Deviation |
|---|---|---|---|
| Actual Realized Compliance | First Principal Component of Questions 1a-1g in Screen 13 | 0.12 | 1.68 |
| Appropriateness of violation of safety norms | First Principal Component of Questions 2a-2d in Screen 14 | -0.14 | 1.53 |
| Agreeableness with government performance | First Principal Component of Questions 3a-3b in Screen 14 | -0.05 | 1.35 |
| Female | =1, if gender is Female | 0.27 | 0.44 |
| Age | age in years | 34.19 | 9.15 |
| Education | =0, if highest educational level is up to class 12 =1, if highest education level is bachelor's degree =2, if highest education level is master's degree or above | 1.12 | 0.65 |
| Income | Log of monthly family income (PPP USD) | 8.39 | 1.79 |
| Health | Health condition on a scale of 0-5 [0 if very poor health, 5 if very good health] | 4.12 | 0.75 |
| Perceived effectiveness | Perceived effectiveness of the safety measures being proposed to counter the spread of COVID-19 | 4.30 | 0.92 |

| | Sample Size | | |
|---|---|---|---|
| N | | 334 | |
| N - Stage1 | Sample size from countries with COVID-19 cases less than 100, as on 21st March | 121 | |
| N - Stage2 | Sample size from countries with COVID-19 cases between 100 and 999, both numbers included, as on 21st March | 108 | |
| N - Stage3 | Sample size from countries with COVID-19 cases more than 999, as on 21st March | 105 | |
| Countries | Number of countries represented in our sample | 43 | |

Table A2: Comparing convexity of the actual and the predicted growth path of COVID-19

|  | (1) | (2) | (3) |
|---|---|---|---|
|  | **Early Phase** | | |
|  | Ratio Actual (n= Slope(N4,N5)/Slope(N3,N4)) | Median Ratio Predicted (p =median Slope(P4,P5)/Slope(N3,P4)) | Difference (n-p) |
| All countries | 5.91 | 2.04 | 3.87*** |
| Germany | 8.30 | 2.38 | 5.92*** |
| USA | 10.32 | 3.22 | 7.10*** |
| France | 3.40 | 2.00 | 1.40** |
| Spain | 6.57 | 2.13 | 4.44*** |
|  | **Later Phase** | | |
|  | Median Ratio Actual (n=median Slope(N4,N5)/Slope(N3,N4)) | Median Ratio Predicted (p=median Slope(P4,P5)/Slope(N3,P4)) | Difference (n-p) |
| All countries | 2.90 | 1.60 | 1.30*** |
| Germany | 2.01 | 2.03 | -0.02 |
| USA | 7.30 | 1.86 | 5.44*** |
| France | 2.33 | 1.74 | 0.59 |
| Spain | 2.52 | 1.68 | 0.84** |

† Note: Col (1) reports the slopes of the straight line connecting the actual number of cases. Col (2) reports the ratio of the median slopes of the straight line connecting the predicted number of cases. Col (3) reports the difference between Col (1) and (2). Ranksum tests indicate the difference is significantly different from 0, implying the degree of convexity is significantly higher in the true data generating process than in the median prediction model. * $p<0.10$, ** $p<0.05$, *** $p<0.01$

Table A3: Prediction Bias and Stages of COVID-19 transmission

|  | (1) | (2) | (3) | (4) | (5) | (6) | (7) | (8) |
|---|---|---|---|---|---|---|---|---|
| **A** | Bias43 | | Bias53 | | AverageBias | | OwnBias | |
| Reference group: Stage1 | | | | | | | | |
| Stage2 | 0.12* | 0.12 | 0.06 | 0.05 | 0.07 | 0.09 | 0.46* | 0.54** |
|  | (0.07) | (0.08) | (0.06) | (0.06) | (0.05) | (0.06) | (0.25) | (0.23) |
| Stage3 | 0.27*** | 0.25*** | 0.16*** | 0.14** | 0.22*** | 0.22*** | 0.98*** | 0.77*** |
|  | (0.07) | (0.08) | (0.06) | (0.06) | (0.05) | (0.06) | (0.25) | (0.25) |
| Constant | 0.30*** | -0.05 | 0.40*** | 0.12 | 0.35*** | -0.03 | -0.02 | -0.83 |
|  | (0.05) | (0.27) | (0.04) | (0.20) | (0.04) | (0.20) | (0.17) | (0.81) |
| Observations | 334 | 309 | 334 | 309 | 334 | 309 | 333 | 308 |
| Other controls | No | Yes | No | Yes | No | Yes | No | Yes |
| **B** | | | | | | | | |
| Reference group: Stage2 | | | | | | | | |
| Stage3 | 0.15** | 0.13 | 0.11* | 0.09 | 0.15*** | 0.13** | 0.52** | 0.23 |
|  | (0.07) | (0.08) | (0.06) | (0.06) | (0.06) | (0.06) | (0.26) | (0.26) |
| Constant | 0.42*** | 0.06 | 0.45*** | 0.17 | 0.42*** | 0.05 | 0.45** | -0.29 |
|  | (0.05) | (0.27) | (0.04) | (0.20) | (0.04) | (0.20) | (0.18) | (0.80) |
| Observations | 334 | 309 | 334 | 309 | 334 | 309 | 333 | 308 |
| Other controls | No | Yes | No | Yes | No | Yes | No | Yes |

[†] Note: In Panel A, prediction biases of participants from countries in Stage 2 and 3 are compared with those from Stage 1 countries. Stage 3 dummy is consistently positive and significant indicating the bias is significantly higher in Stage 3 countries than Stage 1 countries. Panel B compares the same between Stage 3 and Stage 2 countries. The biases are significantly higher in Stage 3 relative to Stage 2 for a number of specifications. Control variables for the incentivized biases include age, gender, health condition, education level, income, treatment, log number of COVID-19 cases as of 21st March. For OwnBias, in addition to the above controls, individual's information bias is included. Heterosckedasticty robust standard errors reported in the parentheses. $p < 0.10$, * $p<0.10$, ** $p<0.05$, *** $p<0.01$.

Table A4: Compliance Indices and Biases

| VARIABLES | (1) Actual Realized Compliance | (2) Actual Realized Compliance | (3) Appropriateness of violation of safety norms | (4) Appropriateness of violation of safety norms | (5) Agreeableness with government performance | (6) Agreeableness with government performance |
|---|---|---|---|---|---|---|
| Bias43 | -0.26 | -0.32# | 0.51*** | 0.40* | 0.28* | 0.31* |
|  | (0.20) | (0.20) | (0.19) | (0.22) | (0.17) | (0.17) |
| Constant | 0.21** | -3.47*** | -0.30*** | 0.20 | -0.14 | -2.96*** |
|  | (0.10) | (1.06) | (0.10) | (0.82) | (0.09) | (0.60) |
| Observations | 334 | 309 | 334 | 309 | 334 | 309 |
| R-squared | 0.005 | 0.133 | 0.024 | 0.052 | 0.010 | 0.189 |
| Bias53 | -0.40* | -0.40* | 0.52** | 0.43* | 0.44** | 0.53*** |
|  | (0.23) | (0.23) | (0.22) | (0.24) | (0.19) | (0.19) |
| Constant | 0.28** | -3.33*** | -0.35*** | 0.05 | -0.22** | -3.15*** |
|  | (0.11) | (1.05) | (0.11) | (0.82) | (0.11) | (0.60) |
| Observations | 334 | 309 | 334 | 309 | 334 | 309 |
| R-squared | 0.008 | 0.133 | 0.017 | 0.049 | 0.015 | 0.200 |
| AverageBias | -0.34 | -0.38* | 0.55*** | 0.44* | 0.37** | 0.43** |
|  | (0.22) | (0.22) | (0.21) | (0.23) | (0.19) | (0.19) |
| Constant | 0.24** | -3.40*** | -0.34*** | 0.12 | -0.18* | -3.04*** |
|  | (0.11) | (1.06) | (0.10) | (0.82) | (0.10) | (0.60) |
| Observations | 334 | 309 | 334 | 309 | 334 | 309 |
| R-squared | 0.007 | 0.134 | 0.022 | 0.052 | 0.013 | 0.194 |
| OwnBias | -0.08* | -0.09** | 0.06** | 0.05 | 0.05** | 0.04 |
|  | (0.04) | (0.04) | (0.03) | (0.04) | (0.02) | (0.03) |
| Constant | 0.19* | -3.68*** | -0.21*** | 0.38 | -0.10 | -2.88*** |
|  | (0.10) | (1.02) | (0.08) | (0.82) | (0.08) | (0.60) |
| Observations | 333 | 308 | 333 | 308 | 333 | 308 |
| R-squared | 0.021 | 0.158 | 0.015 | 0.043 | 0.014 | 0.182 |
| Other controls | No | Yes | No | Yes | No | Yes |

† This table reports results from a regression model where three compliance indices, Actual realized compliance, Appropriateness of violation of safety norms and Agreeableness with government performance, are regressed on prediction bias. Panels A, B, C and D present the coefficient of Bias43, Bias53, AverageBias and OwnBias, respectively. Control variables for the incentivized biases include age, gender, health condition, education level, income, treatment, log number of COVID-19 cases as of 21st March and individual's perception of the effectiveness of WHO recommended safety practices. For OwnBias, in addition to the above controls, individual's information bias is included. Heteroskedasticity robust standard errors reported in the parentheses. # p=0.11, * p<0.10, ** p<0.05, *** p<0.01.

Table A5: Effect of the treatment on bias

| VARIABLES | (1) Bias43 | (2) Bias43 | (3) Bias53 | (4) Bias53 | (5) AverageBias | (6) AverageBias |
|---|---|---|---|---|---|---|
| Treatment=Numerical | -0.12** | -0.17*** | -0.11** | -0.13*** | -0.11** | -0.18*** |
| | (0.06) | (0.06) | (0.05) | (0.05) | (0.05) | (0.05) |
| Constant | 0.47*** | -0.13 | 0.51*** | 0.08 | 0.48*** | -0.03 |
| | (0.04) | (0.27) | (0.03) | (0.21) | (0.03) | (0.22) |
| Observations | 334 | 309 | 334 | 309 | 334 | 309 |
| Other controls | No | Yes | No | Yes | No | Yes |

[†] Note: The table reports the effect of the numerical treatment on prediction bias. The treatment effect is negative and significant, indicating that the prediction bias is substantially lower when the data is represented in terms of number relative to a graph. Control variables for the incentivized biases include age, gender, health condition, education level, log income, log number of COVID-19 cases as of 21st March. Heterosckedasticty robust standard errors reported in the parentheses. * $p<0.10$, ** $p<0.05$, *** $p<0.01$.

# Appendix 2 (Experimental Instructions)

# Experimental Instructions

## Screen 1

### Informed Consent Form

You are being invited to participate in this study conducted Joydeep Bhattacharya (Iowa State University, USA). Your consent is specifically sought on the last page of this form.

### Title of Study

Behavioral Responses to COVID-19

### Purpose

The purpose of this research is to study human responses and coping strategies related to the spread of the Coronavirus. The survey is designed to generate valuable information about your own behavior and perceptions. Besides, it will collect some basic demographic information. Our purpose to collect *aggregated* information about human responses so we can better understand *average* behavior in a population.

### Procedure

In the first part of the study, you will be given some data and asked to make predictions about the future. Following this, you will be asked to answer a short questionnaire on your behavior and perceptions and of those around you related to the spread of coronavirus. We expect you to take no more than 5-6 minutes to complete the entire survey.

### Risks

There are minimal risks of participation in the study. The choices you make will be recorded using a code number, with no link to your identity and *no one* will get to know what you have chosen. The data will be completely anonymized and safely stored on a GDPR-compliant secure cloud server. At no point ever will anyone (including the researchers) be able to connect you, the person, with any of your responses.

### Future Use of Data

The data will be accessed only by the researchers. De-identified information collected about you during this study may be shared with other researchers or used for future research studies. We will not obtain additional informed consent from you before sharing the de-identified data.

### Benefits

You will be paid for participating in this study provided you complete all the questions relating to predictions. For that, you will be paid a fixed amount of 80 cents. You may earn up to an extra 25 cents depending on the accuracy of your predictions (if they are no more than 5% away from the actuals). Feel free to ignore any of the demographic questions you are uncomfortable with. Failure to answer questions of a demographic nature will not affect your base participation payment of 80 cents. All payments will be administered through Amazon's M-Turk.

### Rights:

Irrespective of where you reside, you have

- The right to request access to any personal data held by us.
- The right to erasure of personal data
- The right to restrict processing of your personal data in certain situations.
- The right to withdraw consent

To the extent your personal data are protected by the **GDPR (General Data Protection Regulation)**, you have the right to—

· Know what data we are collecting, where they will be processed, and how they will be used.

· View and correct your personal data.

· Obtain and transfer your personal data to another organization.

· Have certain personal data destroyed (except when data retention is otherwise required or authorized under the GDPR or other controlling legal authority).

· Withdraw your consent to the continued collection of your personal data; and

· Certain other actions described in Chapter III of the GDPR.

**Voluntary Participation and Rights**

Your participation in this research is completely voluntary. You are free to withdraw from the study at any time. All you need to do is close your browser. If you have any questions about the rights of research subjects or research-related injury, please contact the IRB Administrator, 515-294-4566, IRB@iastate.edu, or Director, 515-294-3115, Office for Responsible Research, Iowa State University, Ames, Iowa 50011.

**Questions**

You are encouraged to ask questions at any time during this study. For further information about the study, contact us over email at [joydeep@iastate.edu](mailto:joydeep@iastate.edu).

**Consent and Data Rights Under the European General Data Protection Regulation (GDPR)**

Your participation in this research will involve the collection and processing of your personal data, as described above. Please indicate whether you consent to the collection and processing of your personal data by **placing your initials** underneath the appropriate selection.

____Yes, I consent to the collection and processing of my personal data.

____No, I do not consent to the collection and processing of personal data.

(Your consent is entirely voluntary but declining to provide it may materially impede your ability to participate in this research project.)

**Consent and Authorization Provisions: Consent:** Choose one of the two boxes

Below are two boxes. By clicking the box on the left, you acknowledge, that you voluntarily agree to participate in this study, that the study has been explained to you, that you have been given the time to read the document, and that your questions have been satisfactorily answered. You may print a copy of this informed consent document for your records.

If you don't agree with this consent document, click the box on the right and you will be exited from the study.

| I have read the above information. I have been given an opportunity to ask questions and my questions have been answered to my satisfaction. I agree to participate in this research. My signature below indicates that I have read and understood the description of the study and I agree to participate." | I am not willing to participate in this study. |

## Screen 2

Select the country you currently lived in.

- France
- Germany
- US
- Spain
- Australia
- Brazil
- Bangladesh
- China
- Greece
- India
- Italy
- Mexico
- Nigeria
- Philippines
- Romania
- Russia
- Turkey
- Venezuela
- UAE
- UK
- Other

Mention your country if you have chosen **Other.**

## Screen 3

Thank you for agreeing to participate in our study. The study consists of **four prediction tasks (Task 1, Task 2, Task 3 and Task 4)** explained in detail below and a short demographic survey. You will receive a fixed amount of 50 cents for completing the entire HIT today. In addition, two of the four tasks you complete will be randomly picked and your earnings on these two tasks, if any, will be paid to you. Click next to proceed to the first task.

## Screen 4

**Task 1**

In this part, we will ask you to predict the number of Coronavirus cases that are likely to be recorded in the near future for a certain country.

A certain country, **W,** has recorded a certain number of Coronavirus cases in each of three successive weeks -- call them Week 1, Week 2, and Week 3. We will show you these numbers and **your task will be to predict the number of cases you think will be recorded in this country in Weeks 4 and 5.** We, the researchers, already know the true number of cases in Weeks 4 and 5 in this country. We want to see how close your predictions are to those true values.

Please note: The numbers you will see are actual numbers for a real but unnamed country.

You are asked to make two predictions, one each for Weeks 4 and 5. We will randomly select one of your predictions. **If your prediction turns out to be within 5% of the actual number of cases and if your response is recorded (if you click next) within 60 seconds of starting the task, you will receive a bonus of 25 cents for this task.**

Click next to see the actual numbers for Week 1-3 and type in your predictions for Week 4-5.

---

### Screen 5

**Task 1 (continued)**
A certain country **W** has recorded the following number of Coronavirus cases.

Actual number of cases in Country W in Week 1:  <x>
Actual number of cases in Country W in Week 2:  <x>
Actual number of cases in Country W in Week 3:  <x>

**Now, think carefully and type your predictions. You can win a bonus of 25 cents if you click next within 60 seconds and if your prediction is within 5% of the true number.**

Your prediction on the number of cases in Country W in Week 4: _____
Your prediction on the number of cases in Country W in Week 5: _____

---

### Screen 5 [in the graph treatment]

**Task 1 (continued)**
A certain country **W** has recorded the following number of Coronavirus cases.

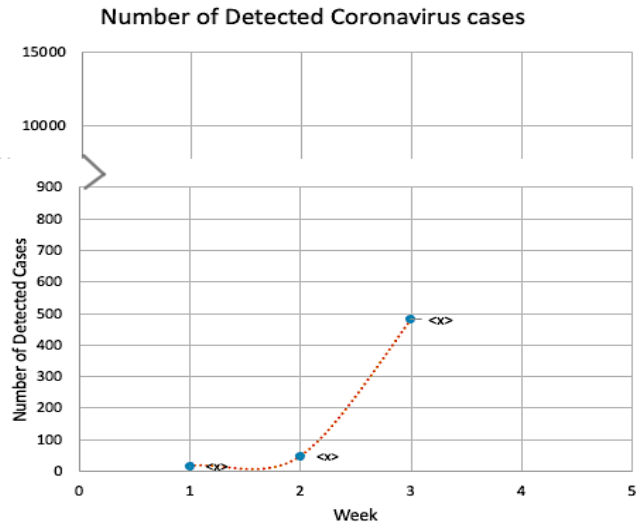

Now, think carefully and type your predictions. You can win a bonus of 25 cents if you click next within 60 seconds and if your prediction is within 5% of the true number.

Your prediction on the number of cases in Country W in Week 4: _____
Your prediction on the number of cases in Country W in Week 5: _____

---

### Screen 6

Thank you for your response to Task 1. For Task 2, please read below.

### Task 2

In this part, we will ask you to predict the number of Coronavirus cases that are likely to be recorded in the near future for a certain country.

A certain country, **X,** has recorded a certain number of Coronavirus cases in each of three successive weeks -- call them Week 1, Week 2, and Week 3. We will show you these numbers and your task will be to predict the number of cases you think will be recorded in this country in Weeks 4 and 5. We, the researchers, already know the true number of cases in Weeks 4 and 5 in this country. We want to see how close your predictions are to those true values.

Please note: The numbers you will see are actual numbers for a real but unnamed country.

You have to make two predictions, one each for Week 4 and 5. We will randomly select one of the predictions. **If your prediction is within 5% of the actual number of cases and if your response is recorded (if you click next) within 60 seconds, you will receive a bonus of 25 cents for this task.**

Click next to see the actual numbers and type in your predictions.

---

### Screen 7

### Task 2 (continued)
A certain country **X** has recorded the following number of Coronavirus cases.

Actual number of cases in Country X in Week 1: <x>
Actual number of cases in Country X in Week 2: <x>
Actual number of cases in Country X in Week 3: <x>

Now, think carefully and type your predictions. You can win a bonus of 25 cents if you click next within 60 seconds and if your prediction is within 5% of the true number.

Your prediction about the number of cases in Country X in Week 4: _____
Your prediction about the number of cases in Country X in Week 5: _____

---

Screen 8

Thank you for your response to Task 2. For Task 3, please read below.

Task 3
In this part, we will ask you to predict the number of Coronavirus cases that are likely to be recorded in the near future for a certain country.

A certain country, **Y,** has recorded a certain number of Coronavirus cases in each of three successive weeks -- call them Week 1, Week 2, and Week 3. We will show you these numbers and your task will be to predict the number of cases you think will be recorded in this country in Weeks 4 and 5. We, the researchers, already know the true number of cases in Weeks 4 and 5 in this country. We want to see how close your predictions are to those true values.

Please note: The numbers you will see are actual numbers for a real but unnamed country.

You have to make two predictions, one each for Week 4 and 5. We will randomly select one of the predictions. If your prediction is within **5% of the actual number** of cases and if your response is recorded (if you click next) within **60 seconds**, you will receive a bonus of 25 cents.

Click next to see the actual numbers and type in your predictions.

---

Screen 9

Task 3 (continued)
A certain country **Y** has recorded the following number of Coronavirus cases.

Actual number of cases in Country Y in Week 1: <x>
Actual number of cases in Country Y in Week 2: <x>
Actual number of cases in Country Y in Week 3: <x>

Now, think carefully and type your predictions. You can win a bonus of 25 cents if you click next within 60 seconds and if your prediction is within 5% of the true number.

Your prediction about the number of cases in Country Y in Week 4: _____
Your prediction about the number of cases in Country Y in Week 5: _____

---

Screen 10

Thank you for your response to Task 3. For Task 4, please read below.

Task 4

In this part, we will ask you to predict the number of Coronavirus cases that are likely to be recorded in the near future for a certain country.

A certain country, **Z,** has recorded a certain number of Coronavirus cases in each of three successive weeks -- call them Week 1, Week 2, and Week 3. We will show you these numbers and your task will be to predict the number of cases you think will be recorded in this country in Weeks 4 and 5. We, the researchers, already know the true number of cases in Weeks 4 and 5 in this country. We want to see how close your predictions are to those true values.

Please note: The numbers you will see are actual numbers for a real but unnamed country.

You have to make two predictions, one each for Week 4 and 5. We will randomly select one of the predictions. If your prediction is within **5% of the actual number** of cases and if your response is recorded (if you click next) within **60 seconds**, you will receive a bonus of 25 cents.

Click next to see the actual numbers and type in your predictions.

---

### Screen 11

### Task 4 (continued)
A certain country **Z** has recorded the following number of Coronavirus cases.

Actual number of cases in Country Z in Week 1:  <x>
Actual number of cases in Country Z in Week 2:  <x>
Actual number of cases in Country Z in Week 3:  <x>

Now, think carefully and type your predictions. You can win a bonus of 25 cents if you click next within 60 seconds and if your prediction is within 1% of the true number.

Your prediction about the number of cases in Country Z in Week 4: _____
Your prediction about the number of cases in Country Z in Week 5: _____

---

### Screen 12

Thank you for your response to Task 4.

What is your rough estimate about the number of people actually **detected (i.e. tested positive)** with coronavirus **in your country**?

What is your guess about the number of people actually **infected (i.e. those who are carrying the virus but have not been tested)** with coronavirus in your country? We expect you will not look up the internet to report your guess.

What is your prediction about the number of people who will be actually **detected (i.e. tested positive)** with coronavirus in your country **one week from now**?

Please click next to fill out a short survey.

---

### Screen 13
### Survey-1

*[Actual realized compliance: 1a – 1g]*

1. To what extent do the following statements describe your actions and decisions in the past one week?

a.   I did not join social gatherings:      1 (Does not describe at all) …. 5 (Describes very well)
b.   I kept a meter's distance with people I had to interact:
c.   I avoided hand shake and other kinds of contacts:
d.   I tried as far as possible to stay at home:
e.   I used hand face mask:
f.   I used hand sanitizers:
g.   I washed my hands frequently:

---

## Screen 14

### Survey-2

*[Appropriateness of safety norms: 2a – 2d]*

2. According to you, indicate how appropriate, the following behavior is in your country, given the current situation related to coronavirus:

a.   People participating at social gatherings, at weddings, pubs, conferences etc. 1 (Very inappropriate) …. 5 (Very appropriate)
b.   People going out to work:
c.   People going out to malls, movie theatres etc.:
d.   People going out to supermarkets, groceries, gas stations, banks etc.:

*[Agreeableness with government performance: 3a – 3b]*
3. How much do you to agree to the following statements:

a. You trust that the government in your country is doing everything possible to address the situation related to coronavirus.
b. Government in your country has been largely truthful about the reports on coronavirus outbreak.
c. Government's imposition of a general curfew in your country (with the exception of essential services).

4. Social distancing measures (e.g. self quarantine and curfew) are effective strategies to slow down the spread of the coronavirus.

---

## Screen 15

### Survey-3
5. Gender: *Male  *Female  *Other
6. Date of Birth: [dropdown]
7. Marital Status: *Married (or has partners)  *Single (or divorced)
8. Religious identity: *Christian  *Muslim  *Hindu  * Buddhist  *Others  *No Religion  *Prefer not to say
9. Number of family members:
10. Monthly family income before tax (in your home country currency):
11. Highest educational levels: *Class X   * Class XII   *Bachelor degree * Master degree and above

12. Do you have health insurance? *Yes   *No
13. Does your family (other than you) have health insurance? *Yes *No
14. On a scale of 1 to 5, how healthy do you think you are? 1(very poor health) … 5(very good health)
15. Do you have the following health conditions? Heart diseases, diabetes, high blood pressure, high lesterol, kidney diseases, cancer related diseases
16. 28. Have you or the major breadwinner of your family lost his/her job in the past one month due to the coronavirus related economic situation? *Yes  *No.